\documentstyle[epsf,twocolumn]{mn}
\newcommand{\etal}{{\sl et~al.}}

\newcommand{\exo}{{\sl EXOSAT}}
\newcommand{\eu}{{\sl EUVE}}
\newcommand{\iu}{{\sl IUE}}

\newcommand{\cms}{\mbox{$\rm cm^{-2}\:$}}

\newcommand{\hi}{\mbox{$\rm {H\,{\sc i}}\:$}}
\newcommand{\hei}{\mbox{$\rm {He\,{\sc i}}\:$}}
\newcommand{\heii}{\mbox{$\rm {He\,{\sc ii}}\:$}}

\newcommand{\tlus}{\mbox{$\rm {\sc tlusty}\:$}}
\newcommand{\iraf}{\mbox{$\rm {\sc iraf}\:$}}

\newcommand{\xsp}{\mbox{$\rm {\sc xspec}\:$}}
\newcommand{\teff}{\mbox{$T_{eff}\:$}}

\title{Evidence for the stratification of Fe in the photosphere of G191$-$B2B}

\author[M.A. Barstow et al.]{
M.A. Barstow$^1$, I. Hubeny$^2$ and J.B. Holberg$^3$
\\
$^1$ {\it Department of Physics and Astronomy, University of Leicester,
University Road, Leicester LE1 7RH, UK}\\
$^2$ {\it Laboratory for Astronomy and Solar Physics, NASA/GSFC, Greenbelt,
Maryland, MD 20711 USA}\\
$^3$ {\it Lunar and Planetary Laboratory, University of Arizona, Tucson, 
AZ 85721, USA}\\
}

\begin{document}

\label{firstpage}

\maketitle

\begin{abstract}

The presence of heavy elements in the atmospheres of the hottest H-rich DA white 
dwarfs has been the subject of considerable interest. While theoretical 
calculations can demonstrate that radiative forces, counteracting the effects of 
gravitational settling, can explain the detections of individual species, the 
predicted abundances do not accord well with observation. However, accurate 
abundance measurements can only be based on a thorough understanding of the 
physical structure of the white dwarf photospheres, which has proved elusive. 
Recently, the availability of new non-LTE model atmospheres with improved atomic 
data has allowed self-consistent analysis of the EUV, far UV and optical spectra 
of the prototypical object G191$-$B2B. Even so, the predicted and observed
stellar fluxes remain in serious disagreement at the shortest wavelengths (below 
$\approx 190$\AA ), while the inferred abundances remain largely unaltered. We 
show here
that the complete spectrum of G191$-$B2B can be explained by a model atmosphere 
where Fe is stratified, with increasing abundance at greater depth. This 
abundance profile may explain the difficulties in matching
observed photospheric abundances, usually obtained by analyses
using homogeneous model atmospheres, to the detailed radiative levitation 
predictions. Particularly as the latter are only strictly valid for
regions deeper than where the EUV/far UV lines and continua are formed. 
Furthermore, the relative depletion of Fe in the outer layers of the atmosphere 
may be evidence for radiatively driven mass loss in G191$-$B2B.

\end{abstract}


\section{Introduction}

Following the first discovery of the presence of heavy elements in the 
photospheres of H-rich (DA) white dwarfs with \iu \ (Bruhweiler \& \ Kondo 
1981, 1983), it is now well established that they are ubiquitous 
in the group of hottest objects, with effective temperatures in excess of 55000K 
(e.g. Marsh \etal \ 1997b; Barstow \etal \ 1993). Extensive further studies in 
the far UV have eventually revealed the presence of absorption lines from C, N, 
O, Si, S, P, Fe and Ni in a number of objects (e.g. Vennes \etal \ 1992; Sion 
\etal \ 1992; Holberg \etal \ 1994; Vennes \etal \ 1996).

Apart from the detectable far UV absorption lines, these heavy elements have 
important effects on other spectral regions. For example, they systematically 
alter the flux level and shape of the optical Balmer line profiles, from which 
\teff \ and log g can be determined. The effect is to lower the measured value 
of \teff \ by several thousand degrees, compared to that determined under the 
assumption that the star has a pure H envelope (Barstow \etal \ 1998). In the 
extreme ultraviolet (EUV), the heavy element opacity dramatically blocks the 
emergent flux, yielding a much steeper short wavelength cutoff than is seen in a 
star with a pure H atmosphere. While, the general shape of the EUV spectrum of 
metal-containing hot DA white dwarfs has been understood qualitatively, since 
Vennes \etal \ (1988) first successfully interpreted the \exo \ spectrum of 
Feige 24 with an arbitrary mixture of elements, quantitative agreement between 
the observations and the predictions of theoretical model atmospheres has proved 
much more elusive. For example, initial attempts failed to match either the flux 
level or even the general shape of the continuum of one of the best studied 
white dwarfs, G191$-$B2B (e.g. Barstow \etal \ 1996). This was eventually 
perceived to be due to inclusion of an insufficiently large number of heavy element 
lines (mainly Fe and Ni). Improved non-LTE calculations, containing some
9 million predicted Fe and Ni lines, rather than just the 300,000 or so observed 
experimentally, were able to provide a self-consistent model which could 
accurately
reproduce the EUV, UV and optical spectra (Lanz \etal \ 1996). Subsequently, 
other authors have obtained similar
results with both LTE and non-LTE models (e.g. Wolff \etal \ 1998; Koester \etal 
\ 1997; Chayer \etal \ 1997). Interestingly, Wolff \etal \ (1998) obtained
a lower Fe abundance from their far UV analysis than did Lanz \etal \ (1996;
$Fe/H=2\times 10^{-6}$ and $1\times 10^{-5}$ respectively), while agreeing
with the amount of Fe ($1\times 10^{-5}$) required by the EUV spectra.
However, we note that Wolff \etal \ (1998) only study the restricted wavelength
range available from the GHRS spectra, confining their analysis to only a few, 
relatively weak FeV lines. In contrast, Lanz \etal \ (1996) considered the best
match to all the Fe lines visible in the \iu \ spectrum. Furthermore, the
far UV lines are much less sensitive to the Fe abundance than the EUV continuum.
Consequently, it is likely that the apparent discrepancy is not significant.

Despite these advances important problems remain. First, the good agreement 
between the observed EUV spectrum of G191$-$B2B and latest model predictions can 
only be obtained by inclusion of a significant quantity of helium, either in the 
photosphere or in the form of an interstellar/circumstellar \heii \ component 
(see Lanz \etal \ 1996). At the comparatively limited  $\approx 0.5$\AA \ 
resolution of \eu \ in the region of the \heii \ Lyman series, and the 
consequent blending of these lines
with features from heavier elements, the inferred He contribution cannot be 
directly detected. If He is really present, two alternative interpretations 
arise. Either there is an interstellar/circumstellar component or
the material resides predominantly in the stellar photosphere. If the former 
explanation holds, the amount
of \heii \ required implies an extremely high ionization fraction (80\% ) when 
compared with the measured \hei \ column density. This is a much higher value 
than appears to be typical of the local interstellar medium (ISM) in general 
(Barstow \etal \ 1997a). On the other hand, if there is a significant 
photospheric component the implied
abundance of He/H=$5.5\times 10^{-5}$ is in disagreement
with the upper limit of $2\times 10^{-5}$ imposed by the
absence of a detectable 1640\AA \ feature in the UV.

A partially successful attempt has been made to resolve these issues by adopting 
a physically more realistic
model, where the helium is gravitationally stratified rather than homogeneously 
distributed within the atmosphere (Barstow \& \ Hubeny 1998). The required 
interstellar He ionization fraction remains high
(59\% ), compared to that of the local
ISM, but the predicted strength of the 1640\AA \ \heii \ line becomes consistent 
with observation. Unfortunately,
with the stratified model, the predicted \heii \ Lyman series lines are somewhat 
stronger than can be accommodated by the \eu \ spectrum. In the end, the issue
of the \heii \ component will only be solved when a much higher resolution 
spectrum is obtained, capable of resolving any \heii \ lines from those of 
heavier elements. The J-PEX sounding rocket spectrometer, with an expected 
flight in early 1999, should provide such data (Bannister \etal \ 1999).

The second major difficulty in understanding the spectrum of G191$-$B2B has 
received considerably less attention.
Indeed, to some extent, with the recent successes in dealing with the optical, 
UV and medium-long wavelength
\eu \ simultaneously, it has been specifically ignored.
All the results of Lanz \etal \ 1996, Barstow \& \ Hubeny (1998) and others, 
discussed above, only consider the EUV spectral data longward of $\approx 
180$\AA . Nevertheless, there is significant flux detected shortward of this, in
the \eu \ short wavelength channel and in the soft X-ray.
Furthermore, the flux level predicted by the most successful current models is 
between five and ten times
that observed.

A complete understanding of the atmosphere of G191$-$B2B requires that we also 
explain the short wavelength spectrum as well as the longer wavelength spectral 
ranges.
We show here that the {\it entire} spectrum of G191$-$B2B,
from the short wavelength EUV to the optical,
can be explained if the Fe known to be present in the atmosphere is not 
homogeneously mixed but stratified, with a decreasing abundance towards the 
outer layers of the stellar envelope. Such a heavy element distribution may
be an indication of ongoing mass loss from the star, which has important 
consequence for the spectral evolution of
G191$-$B2B and hot DA white dwarfs in general.

\section{EXAMINATION OF THE SHORT WAVELENGTH FLUX PROBLEM}

\subsection{Observations}

As in our earlier papers on G191$-$B2B, we utilised the `dithered' \eu \ 
spectrum obtained on 1993 December 7-8
(see Lanz \etal \ 1996) and the coadded \iu \ echelle spectrum of Holberg \etal 
\ (1994). The \eu \ `dither' mode consists of a series of pointings
slightly offset in different directions from the nominal source position
to average out flat field variations.
As these data have been extensively described elsewhere, we just give 
a brief
summary here. The \eu \ exposure times were 58,815s, 49239s and 60,816s in the 
SW, MW and LW ranges respectively. We assume that the residual efficiency 
variation, after the effect of the `dither', is 5\% , as observed for HZ43 (e.g. 
Barstow Holberg and Koester 1995; Dupuis \etal \ 1995), quadratically adding a systematic 
error of this magnitude to the statistical errors of the data. As one of the 
brightest EUV sources in the sky, the spectrum of G191$-$B2B is well-exposed 
across most of the wavelength
range, achieving the maximum signal-to-noise possible with \eu \ (limited by the 
residual fixed pattern efficiency variation) at the maximum spectral resolution. 
Since the raw spectra oversample the true resolution by a factor $4$, we have 
generally chosen to bin the data by this factor during the spectral analysis. 
However, in the SW range, the heavy element opacity causes a dramatic reduction 
in the observed stellar brightness. Consequently, in an exposure optimised for 
the MW and LW ranges, the signal-to-noise at shorter wavelengths is
very low. To approach the signal-to-noise achieved at longer wavelengths and 
produce a data set with no bins containing zero counts, it was necessary to 
rebin the data below $\approx 190$\AA \ by a further factor 8 (32 in total). 
Inevitably, detailed spectral line information is lost but the data can 
otherwise constrain the model spectra. Figure 1 shows the complete \eu \ 
spectrum of
G191$-$B2B, comparing it to best fit model of Lanz \etal \ 
(1996).

Recent work (Barstow Hubeny \& \ Holberg 1998) has shown that the value of \teff 
\ determined from an analysis of the optical Balmer lines is sensitive to a 
combination of non-LTE effects and heavy element line blanketing. In their 
analysis, Lanz \etal \ (1996) adopted an effective temperature of 56000K, after 
taking these effects into account. However, their work was carried out using a 
model
grid with a fixed value of the surface gravity (log g=7.5). The more complete 
analysis of Barstow \etal \ (1998) spans the log g range from 7.0 to 8.0 and 
yields
a slightly lower temperature of 53720K, from a combined
Balmer and Lyman line analysis. For this study we adopt \teff =54000K and log 
g=7.5 in the model calculations.

\subsection{Non-LTE spectral models with heavy elements}

The homogeneous non-LTE heavy element rich calculations
used here originate in the work of Lanz \etal \ (1996) and Barstow \etal \ 
(1998) and have been described extensively in those papers. Briefly, the models 
include a total of 26 ions of H, He, C, N, O, Si, Fe and Ni in calculations with 
the programme \tlus  (Hubeny 1988; Hubeny \& \ Lanz 1992, 1995). Radiative data 
for the light elements have been extracted from TOPBASE, the database of the 
opacity project (Cunto \etal \ 1993), except for extended models of carbon atoms 
(K. Werner, private communication). For iron and nickel, all the levels 
predicted by Kurucz (1988) are included, taking into account the effect of over 
9.4 million lines.

Barstow \etal \ (1998) computed a model grid over a range of \teff \ from 52000K 
to 68000K and log g from 7.0 to 8.0
but only considered a single representative value of the
Fe abundance. These calculations have also been extended
to deal with helium stratification by Barstow \& \ Hubeny
(1998). As part of our continuing study of G191$-$B2B-like
hot DA white dwarfs we have extended the stratified computations to match the 
grid of Barstow \etal \ (1998) in \teff \ and log g while enlarging the range of 
Fe abundances considered (see table~\ref{mgrid}).

\begin{table}
\caption{Physical parameters adopted for model
calculations with a homogeneous mixture of Fe}
\begin{tabular}{ll}
Parameter & Grid point values \\
\teff \ (K)& 52000, 54000, 56000, 63000, 68000\\
log g & 7.0, 7.5, 8.0 \\
log $M_H$ ($M_\odot$) & -14.0, -13.42, -12.92\\
log(Fe/H) & -5.5, -5.0, -4.5 \\
\end{tabular}
\label{mgrid}
\end{table}

In addition to the calculation of the intrinsic stellar EUV spectrum, it is 
necessary to take account of the effect of the intervening interstellar medium. The 
basic model
that deals with \hi , \hei \ and \heii \ opacity is now
well-established (Rumph Bowyer \& \ Vennes 1994) and we
apply the modifications described by Dupuis \etal \ (1995) to treat the 
converging line series near the \hei \ and \heii \ edges. 

\subsection{Spectral analysis}

The analysis technique used to compare models and data
has been described extensively in several earlier papers
(e.g. Lanz \etal \ 1996; Barstow \etal \ 1997a,b etc.).
Hence, we just give a brief resum\'e here. We utilise the programme \xsp \ to 
fold model spectra through the \eu \ instrument response, taking into account 
the overlapping higher spectral orders in the instrumental effective area and 
applying the long wavelength corrections described by Dupuis \etal \ (1995). 
Goodness of fit is determined using a $\chi^2$ statistic and the best agreement 
between model and data is achieved by seeking to minimise the value of that 
parameter. While visually, good agreement can be achieved between model and data 
disagreements in some details of the line strengths often lead to high values of 
the reduced $\chi^2$ ($\chi^2/\nu$, where $\nu $ is the number of degrees of 
freedom). Formally, a good fit should have $\chi^2_{red}\approx 2$ or less and 
if it is much greater than this, estimates on the parameter uncertainties cannot 
be evaluated from the change in $\chi^2$ 
according the usual values of $\Delta \chi^2$ 
(e.g. $\Delta \chi^2=5.89$ for $1\sigma$ uncertainty and 5 degrees of freedom,
see Press \etal \ 1992). 

An alternative way to estimate such uncertainties is to make use of the F test, 
which calculates the significance of differences in the values of $\chi^2$ 
determined from separate fits. The F parameter is simply the ratio of the
two values of $\chi^2_{red}$. The significance of its value depends on the 
number of degrees of freedom and can be determined from standard tables.
This can 
be used to determine whether or not one model might provide a significantly 
better fit than another. Uncertainties are then estimated by tracking the value of 
F as an individual parameter is varied until it reaches a predetermined value 
corresponding to a particular significance.

\subsection{Comparison of models and data}

\begin{figure*}
\leavevmode\epsffile{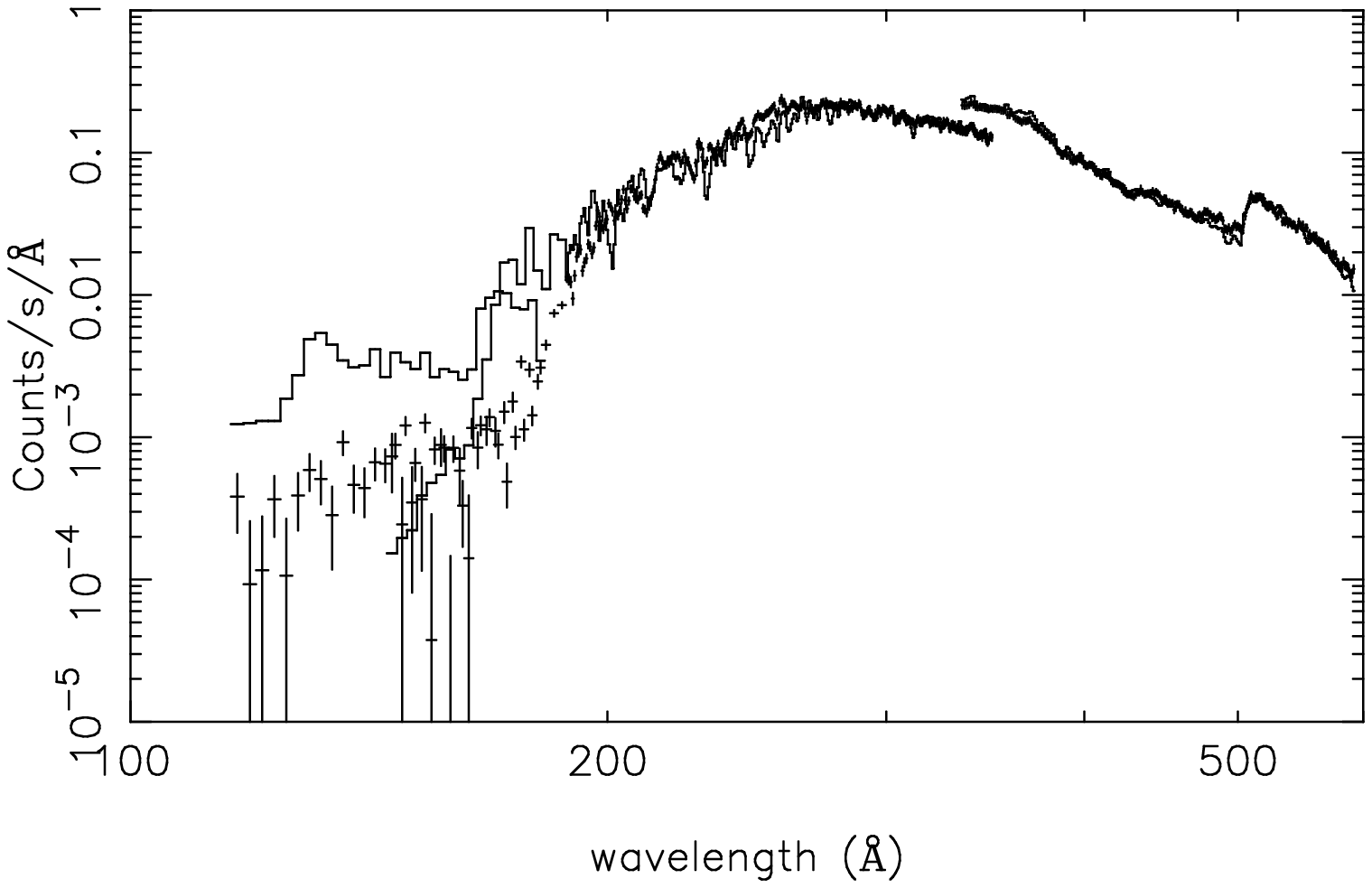}
\caption{Comparison of the complete \eu \ count spectrum of G191$-$B2B 
(error bars) with a single 
model (separate histograms for SW, MW and LW channels running left to right)
that gives the optimum match to the $\lambda > 180$\AA \ range (\hi 
$=2.1\times 10^{18}$,
\hei $=1.9\times 10^{17}$, \heii $=3.5\times 10^{17}$,
$M_H=1.2\times 10^{-13}M_\odot$, Fe/H$=8.5\times 10^{-6}$, log g$=7.5$,
\teff $=54000$K). The discontinuities near 170\AA \ and
320\AA , where the short wave (SW), medium wave (MW) 
and long wave (LW) spectrometer channels overlap, arises from the differing 
spectrometer effective area for which these data are not corrected.}
\label{f1}
\end{figure*}

The work of Lanz \etal \ (1996) and Barstow \& \ Hubeny (1998) has been very 
successful in explaining the spectrum of G191$-$B2B at wavelengths longward of 
$\approx 180$\AA . However, taking the models that best match that spectral
region and extending the comparison to shorter wavelengths
reveals a significant discrepancy, with the observed flux falling well below 
that predicted (figure~\ref{f1}). The largest difference is nearly an order of 
magnitude.
Closer inspection of figure~\ref{f1} reveals that, while the onset of the 
disagreement is quite sudden, at $\approx 190$\AA , the fit is not perfect at 
the longer wavelengths. Agreement is very good above 260\AA , but
there are some significant differences between model and
data between 190\AA \ and 260\AA .

\begin{figure*}
\leavevmode\epsffile{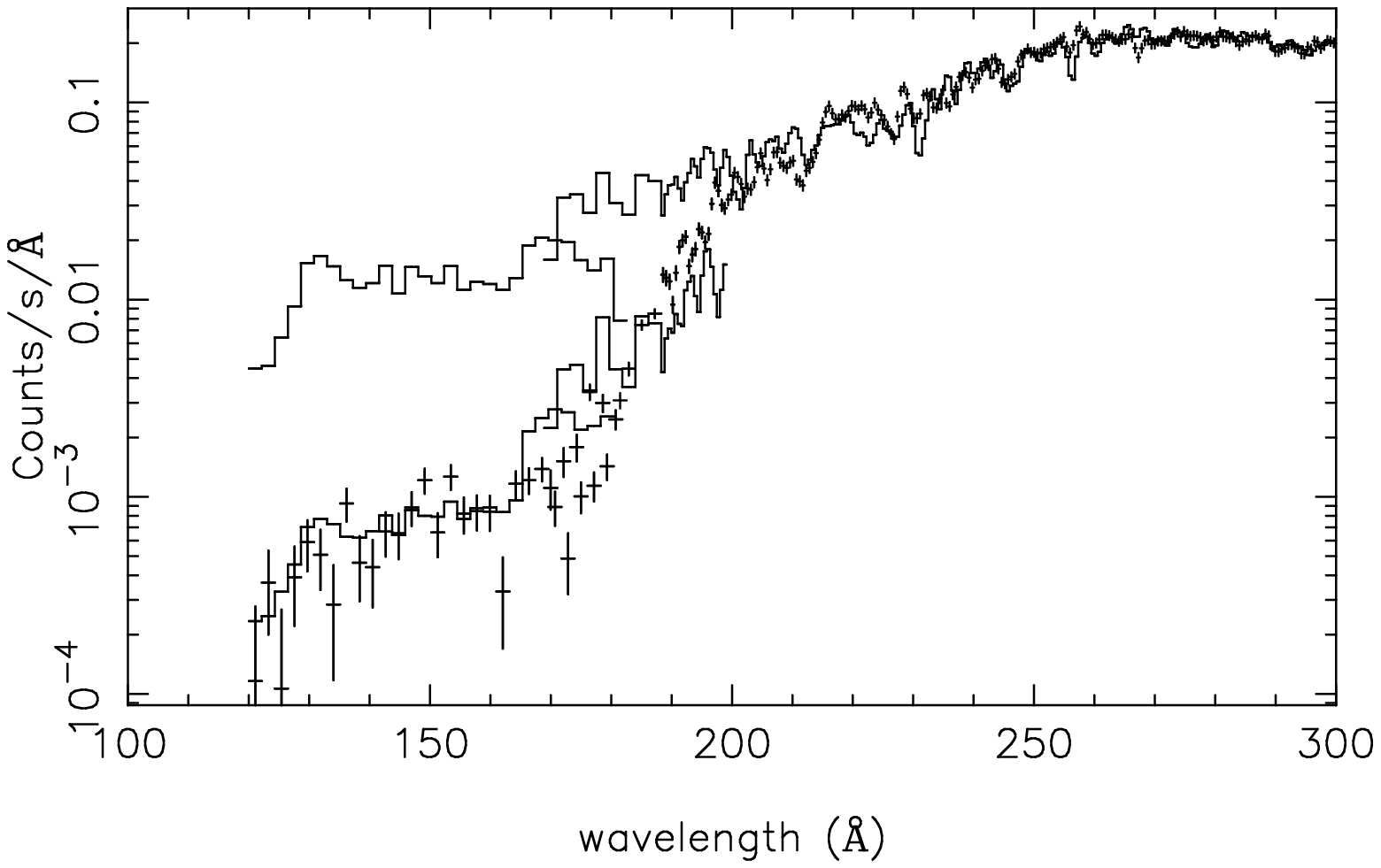}
\caption{Comparison of the  100--300\AA \ region of the \eu \ count spectrum of 
G191$-$B2B with a model (upper pair of histograms; SW range left, MW range right)
that gives the optimum match to the $\lambda > 210$\AA \ 
range (\hi $=2.1\times 10^{18}$,
\hei $=1.7\times 10^{17}$, \heii $=3.6\times 10^{17}$,
$M_H=1.2\times 10^{-13}M_\odot$, Fe/H$=3.8\times 10^{-6}$, log g$=7.5$,
\teff $=54000$K). Also shown is the higher Fe abundance model
(lower pair of histograms, Fe/H$=4\times 10^{-5}$) which best 
matches the  shorter wavelength data
(in the short wavelength region only).
Some of the overlapping MW data points have been removed for clarity.
}
\label{f2}
\end{figure*}

Interestingly, the match of the MW and LW data to the
model can be much improved by restricting further the
wavelength range under consideration. For example, ignoring the data shortward 
of 210\AA \ delivers a large
improvement in the fit, changing the value of $\chi^2$ by more than a factor 2, 
from 4827 (530 degrees of freedom)
to 2064 (491 dof). Applying the F-test to these values
shows that the improvement is hugely significant. Furthermore, this is coupled 
with a reduction in the
required Fe abundance to Fe/H$=3.8\times 10^{-6}$. The quality of 
agreement with the long wavelength data, already very good with the current 
generation of models, remains more or less unchanged. Therefore, nearly all the 
improvement in the fit lies between 210\AA \ 
and $\approx 300$\AA , as illustrated in
figure~\ref{f2}. Here, we highlight the most interesting spectral range, between 
100\AA \ and 300\AA , below which the source is 
barely detectable with \eu . As 
expected from the above discussion, the agreement between model and data is 
excellent above 210\AA , but there is an increasing discrepancy to shorter 
wavelengths.

The difference between the model prediction and observation (see 
figure~\ref{f2}) is characterised mainly by a difference in slope. A clue as to 
the possible explanation of the short wavelength flux discrepancy can be found 
by considering the region below 180\AA \ separately from the rest of the 
spectrum. An improved match to the observed flux level is obtained by allowing 
the Fe abundance to vary freely, also shown in figure~\ref{f2}, after fixing the 
interstellar columns at the values determined from the longer wavelength fit. 
However, this exercise yields a Fe/H ratio of $4\times 10^{-5}$, much higher 
than either the values obtained
in the analyses of Lanz \etal \ (1996) and Barstow \& \ Hubeny (1998) or with 
the more restricted short wavelength limit considered here. 

\begin{figure*}
\leavevmode\epsffile{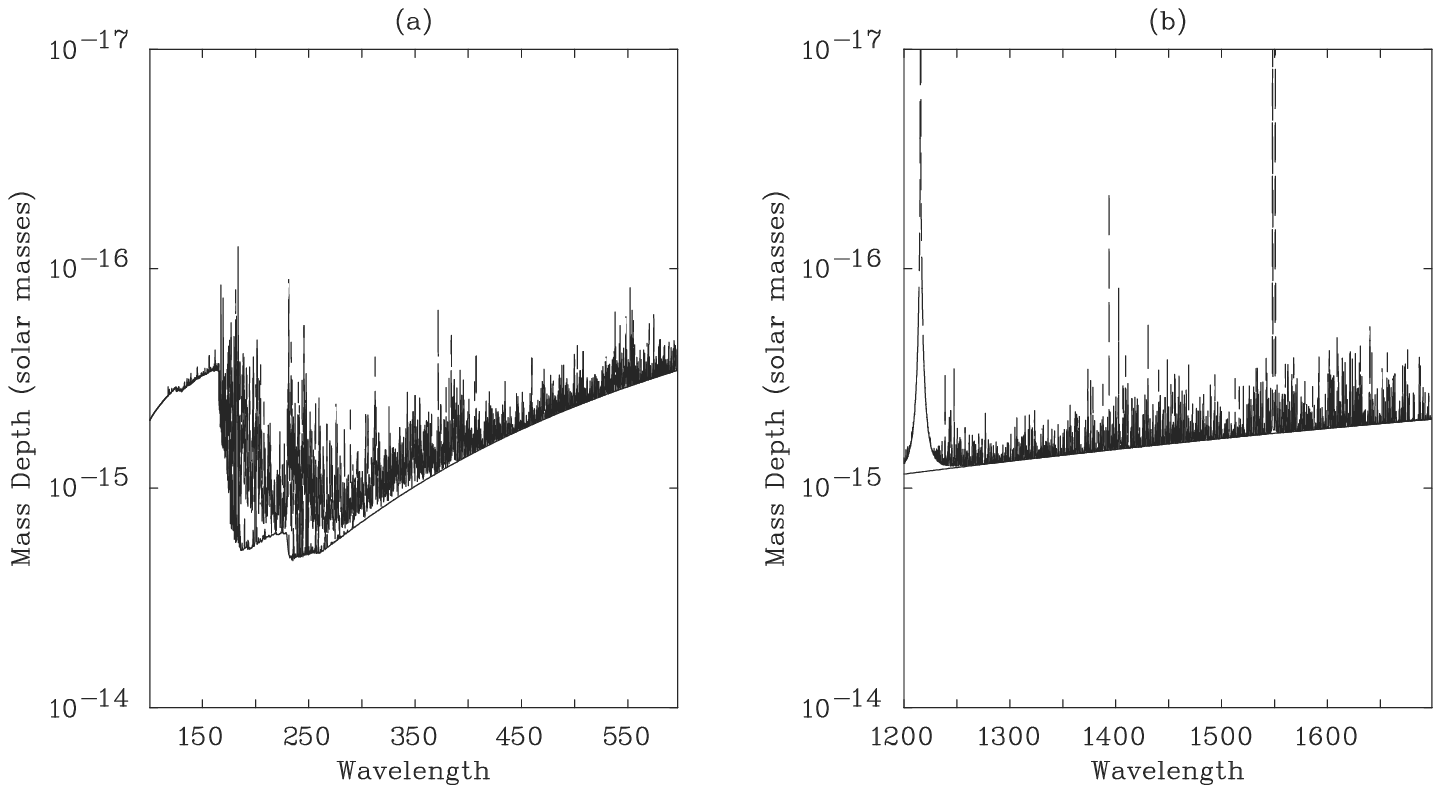}
\caption{Mass depth ($\rm \Delta M$, total mass above the point of 
interest) of the line (upper, dashed curve) and continuum (lower, solid 
curve) formation
at monochromatic optical depth $\tau_\nu=2/3$ as a function of wavelength in the 
EUV and far UV. 
}
\label{f3}
\end{figure*}

Figure~\ref{f3} shows the mass depth ($\rm \Delta M$, total mass above the point of 
interest)
of the EUV line and continuum at
monochromatic optical depth $\tau_\nu=2/3$ as a function of wavelength
computed for the model that gives the best overall fit to the data. It
can be seen that there is a steep change in the depth at which both the
continuum and lines are formed at $\approx 180$\AA , just where short
wavelength discrepancy begins (see figures~\ref{f1} and ~\ref{f2}).
The sharp decrease in the depth of the continuum formation between 190
and 160\AA \ is a combined result of a number of intervening continuum
edges, mostly of FeV ($\lambda $ 165.5, 173.1, 176.1, 180.2, 184.3),
and partly NiV ($\lambda $ 163.2, 171.3, 174.9, 177.0, 180.0). There
are also edges of light ions, namely CIV ground-state (192.2), NIV (160.1,
179.4) and OIV (160.3, 181.1), which are less important than the FeV features.
Coupling this with the larger Fe abundance required to match the short
wavelength data leads us to suggest that the Fe is not homogeneously
mixed in the atmosphere but has a depth dependent abundance. We examine
this possibility in the rest of this paper.

\section{Non-LTE heavy element-rich models with a stratified Fe component}

Barstow \& \ Hubeny (1998) have already reported on the effects of helium 
stratification in heavy element-rich model atmospheres. They utilised the 
programme \tlus \ (Hubeny 1988; Hubeny \& \ Lanz 1992, 1995; Lanz \etal \ 1996) 
including modifications to allow the abundance of any element at any depth point 
within an atmosphere to be a completely free parameter. However, to generate a 
physically realistic model it would be necessary to carry out diffusion 
calculations for every element. In the case of helium, this is comparatively 
straightforward. As pointed out by Vennes \etal \ (1988), the radiation pressure 
on helium is insufficient to counteract the downward force of gravity
to a degree yielding a photospheric helium abundance large enough to
explain the observed EUV/soft X-ray flux deficiency, due to 
the comparatively small number of lines in the EUV waveband. Hence, 
Vennes \etal \ (1988) proposed a model where the the relative 
abundance of He is determined by the equilibrium between ordinary
diffusion and gravitational settling and depends on 
effective temperature, surface gravity and the mass of the overlying H layer. 
Barstow \& \ Hubeny (1998) adopted the approach of Vennes \etal \ (1988) to 
calculate the depth dependent abundance profile of He for their stratified 
models and this is also the case in the models used in this work.

For heavier elements, the number of lines found in the EUV is much greater than 
for helium and radiative forces become important. In these circumstances, simple 
diffusive equilibrium is no longer an adequate treatment of the relative 
abundances of the elements. Detailed models dealing with the effects of radiative 
levitation have been constructed by several workers. Most recently, Chayer, 
Fontaine \& \ Wesmael (1995a) and Chayer \etal \ (1994) have carried out 
calculations that include all the heavy elements so far found in the atmosphere 
of G191$-$B2B.
Unfortunately, for our purposes, the range of mass depth considered by them, 
running from $\rm \Delta M/M\approx 10^{-4}$ upto $\approx 10^{-15}$, correponds 
to a region of the envelope below that of the line/continuum formation depths in 
the \tlus \ models. We note that Chayer \etal \ (1994) consider
the fractional mass depth $\rm \Delta M/M$, whereas we deal with the total mass
$\rm \Delta M$, independent of the mass of the star. However, 
it is a simple matter
to convert our scale to theirs by dividing $\rm \Delta M$ by the known mass of
G191$-$B2B ($\approx 0.5M_\odot$, Marsh \etal \ 1997a).
Hence, with line/continuum formation depths 
above $\rm \Delta M\approx 3\times 10^{-16}-1\times 10^{-15}$
(corresponding to $\rm \Delta M/M\approx 6\times 10^{-16}-2\times 10^{-15}$),
there is little information in the Chayer 
\etal \ (1994, 1995a) work that could be usefully incorporated into the \tlus \ 
models.

Since, we have no a priori information on the possible depth dependent abundance 
of Fe, either from observation or a detailed physical model, we have taken two 
somewhat arbitrary approaches in an attempt to produce one or more
atmosphere models that can match the complete \eu \ spectrum of G191$-$B2B. 
First, we looked at a series of simple `slab' models, where we divide the 
atmosphere into
two or three discrete regions and fix the Fe abundance at a constant value 
within these depth ranges but allow it to vary from region to region. As 
starting points we used the nominal Fe abundances determined from the best fit 
models
to the short wavelength and medium wavelength \eu \ spectra and placed the slab 
divisions near the continuum
formation depth. We consider models with two and three layers, but it
is important to note that 
the resulting discontinuities are unphysical and that this choice is only a rough
attempt to mimic the true depth dependence of the Fe abundance
that might be expected from diffusion equilibrium. 
Table~\ref{slg}  lists the details of the
slab models, giving Fe abundances and depth ranges, the quoted depth 
representing the lower limit of the given abundance for each layer. We note that 
the abundances of all other elements were constant, as specified in Barstow \& \ 
Hubeny (1998;
C/H$=2.0\times 10^{-6}$, N/H$=1.6\times 10^{-7}$, O/H$=9.6\times 10^{-7}$, 
Si/H$=3.0\times 10^{-7}$, Ni/H$=1.0\times 10^{-6}$).

\begin{table*}
\caption{Fe abundances and depth structure of slab models
computed using \tlus . Depths are in units of $M_\odot$.}
\begin{tabular}{lcccccc}
Model index & Depth top zone & Fe/H & Depth 
(optional) mid zone & Fe/H & Depth deep zone & Fe/H \\
fe1 & $6.2\times 10^{-16}$ & $2\times 10^{-5}$ &
      $1.1\times 10^{-15}$ & $1\times 10^{-5}$ &
      $8.3\times 10^{-13}$ & $4\times 10^{-6}$ \\
fe2 & $6.2\times 10^{-16}$ & $2\times 10^{-5}$ &&& 
      $8.3\times 10^{-13}$ & $1\times 10^{-6}$ \\
fe3 & $6.2\times 10^{-16}$ & $3\times 10^{-6}$ &&& 
      $8.3\times 10^{-13}$ & $2\times 10^{-5}$ \\
fe4 & $6.2\times 10^{-16}$ & $3\times 10^{-6}$ &&& 
      $8.3\times 10^{-13}$ & $4\times 10^{-5}$ \\
fe5 & $6.2\times 10^{-16}$ & $1\times 10^{-6}$ &&& 
      $8.3\times 10^{-13}$ & $2\times 10^{-5}$ \\
fe6 & $6.2\times 10^{-16}$ & $1\times 10^{-6}$ &&& 
      $8.3\times 10^{-13}$ & $4\times 10^{-5}$ \\
fe9 & $6.2\times 10^{-16}$ & $5\times 10^{-7}$ &&& 
      $8.3\times 10^{-13}$ & $6\times 10^{-5}$ \\
fe10& $6.2\times 10^{-16}$ & $1\times 10^{-7}$ &&& 
      $8.3\times 10^{-13}$ & $4\times 10^{-5}$ \\
fe11& $6.2\times 10^{-16}$ & $1\times 10^{-7}$ &&& 
      $8.3\times 10^{-13}$ & $1\times 10^{-4}$ \\
fe12& $4.6\times 10^{-16}$ & $1\times 10^{-7}$ &
      $1.4\times 10^{-15}$ & $3\times 10^{-6}$ &
      $8.3\times 10^{-13}$ & $6\times 10^{-5}$ \\
fe13& $4.6\times 10^{-16}$ & $1\times 10^{-7}$ &
      $1.1\times 10^{-15}$ & $3\times 10^{-6}$ &
      $8.3\times 10^{-13}$ & $6\times 10^{-5}$ \\
fe14& $4.6\times 10^{-16}$ & $1\times 10^{-7}$ &
      $8.5\times 10^{-16}$ & $3\times 10^{-6}$ &
      $8.3\times 10^{-13}$ & $6\times 10^{-5}$ \\
\\
Diffusion/levitation & $M_H$ & log $\rm g_{rad}$ & Reservoir Fe/H \\
fe7 & $10^{-14}$ & 7.4 & $2\times 10^{-4}$\\
fe8 & $10^{-15}$ & 7.45 & $2\times 10^{-4}$\\
\end{tabular}
\label{slg}
\end{table*}

A second approach to modelling the Fe abundance was to
modify the diffusion calculation for gravitational
settling of helium to deal with Fe (see Vennes \etal \ 1988; Barstow \& \ Hubeny 
1998). This is straightforward, since the atomic mass and effective charge are 
free parameters in the calulation and can be adjusted. However,
in dealing with helium, it is realistic to assume that there is an infinite 
reservoir overlaid by a thin H shell, a useful boundary condition. As Fe is not 
a product of nuclear burning in a white dwarf of normal mass, such as G191$-
$B2B, it can only exist as a trace element and the 
Fe reservoir cannot be specified in the same way. In this case it is necessary 
to assume that there is some limiting
Fe abundance in the deeper layers of the atmosphere.
Irrespective of any assumptions about this limiting abundance, what is clear 
from a range of test diffusion calulations for Fe is that the rate of change of 
the abundance profile is very steep. That is, for a sensible
abundance of Fe at the continuum formation depth, the outer layers are 
completely depleted of Fe while the Fe abundance in the deeper regions is so 
high that the emergent EUV flux is negligible. This is not too suprising, as the 
atomic mass of Fe is fourteen times that of He, and highlights the already well
documented fact that radiative levitation effects are necessary to explain the 
presence of photospheric Fe.

We have not developed a complete radiative levitation calculation for this work 
but the possible effects can be examined, to first order, by applying a reverse 
acceleration term in the diffusive equilibrium calculation. Examination of 
radiative acceleration
predictions for Fe (see figure 12 of Chayer \etal \ 1995a)
shows that the value of $\rm g_{rad}$ reaches a plateau-like maximum near the 
stellar surface. Consequently, we can adopt a constant value for the radiative 
acceleration term in our calculations. It is important to stress that
our choice of log $\rm g_{rad}$ and the reservoir Fe abundance are entirely 
empirical, based on achieving
Fe abundances that approximately correspond to those included in the slab models 
and which also match the range observed in the homogeneous analysis of section 3 
(see table~\ref{slg}). 

\section{Stratified analysis of the EUV spectrum of G191$-$B2B}

In principle, a detailed study of any individual star requires a grid of models 
to be calculated spanning the possible range of values of all parameters. In 
practice, this is not feasible for a star like G191$-$B2B because of the number 
of free parameters that must be considered.  
A heavy element-rich atmospheric model is specified by  \teff \ and log g,  plus the 
abundance of each element included in addition to hydrogen - seven in the models 
used here. For G191$-$B2B and similar stars, determination of the values of all 
the parameters is a multi-stage process. Temperature and gravity are estimated 
from the Balmer lines while abundances can be determined from an analysis of the 
far UV absorption line strengths. Abundance determinations are typically 
iterative, with initial estimates made using a first guess at the composition 
refined by recalculation of a fully converged model. We now know that, to 
achieve a completely consistent interpretation of the data, it is also necessary to 
redetermine \teff \ and log g in the light of the measured heavy element 
abundances (Barstow \etal \ 1998).

Most analyses performed have made the reasonable simplifying assumption that the 
stellar envelopes are homogeneous. Once it is necessary to consider depth 
dependent abundances, the number of possible variables increases dramatically, 
since it is then necessary to specify the element abundances at each depth 
point. Since, the models constructed for this work have 70 depth points, this 
could mean that it is necessary to deal with $\approx 70$ times the number of 
variables handled in the homogeneous work unless the problem is restricted in 
some
sensible way. The dominant opacity source in the EUV is Fe, therefore, we have 
chosen to confine the analysis to the study of Fe stratification and assume that 
all other heavy elements are homogeneously mixed, with the exception of helium 
which is also stratified as described above and by Barstow \& \ Hubeny (1998).

Even dealing with a single element it is necessary to define the abundance at 70 
depth points and the problem can be reduced further by dividing the atmosphere 
into a smaller number of discrete regions or slabs or trying to specify a smooth 
abundance profile with a smaller number
of diffusion/levitation parameters, as described in section
3 above. Even so, there is no single variable that can be adjusted to yield the 
desired result in terms of matching both short and longer wavelength \eu \ 
spectra at the same time. However, from the discussion of the short wavelength 
problem (section 2), it is at least possible to express the goal as one of 
steepening the short wavelength region
of the spectrum below $\approx 200$\AA \ compared to the
longward flux level. 

Taking as a starting point the Fe abundances determined from separate fits to 
short and long wavelength EUV spectra, the model grid listed in table~\ref{slg} 
has been constructed in an incremental way, in response to the results of a fit 
to an earlier model. In each new model, only one or two small changes were made 
from the previous example, devised to
bring the predicted spectrum into closer agreement with that observed
(but not always successfully). We 
consider the results of all these analyses together in table~\ref{fits}, 
together with the
best-fit homogeneous model, for reference. The probability
that the best fit model (fe6) is a significant improvement on each of the other 
models is calculated and listed in table~\ref{fits}. In each case, the entire 
useful EUV spectrum of G191$-$B2B from 120\AA \ to 600\AA \ was considered. The  
interstellar column densities were allowed to vary completely freely while H 
layer mass (for H and He stratification), \teff \ and log g were fixed at their 
nominal values of $M_H=1.2\times 10^{-13}$, 54000K and 7.5 respectively.

\begin{table*}
\caption{Best fit values of the interstellar columns and the value of $\chi 
^2_{red}$ (with 584 degrees of freedom)for the grid of models with stratified Fe 
abundances. The final column gives the probability that model fe6 is a better 
match to the data than the other model, determined using the F-test. Figures
in brackets ($\chi^2_{red}$ and F test probability) are the results of fitting the
models to two of the strongest FeV lines (1373 and 1376\AA) in the \iu \ spectrum.}
\begin{tabular}{lccccc}
Model index &
\hi \ ($\times 10^{18}$\cms) & \hei \ ($\times 10^{17}$\cms) & \heii \ ($\times 
10^{17}$\cms) & $\chi^2_{red}$ & F test probability (\% ) \\
Homogeneous & 2.15 & 2.18 & 1.18 &21.0 & $>99.99$ \\
fe1  & 2.02 & 1.76& 7.78& 28.2& $>99.99$\\
fe2  & 2.09& 1.84& 5.20& 45.4& $>99.99$\\
fe3  & 1.96& 1.76& 4.92& 15.8 (1.62)& 89.5 (95)\\
fe4  & 2.09& 2.06& 2.21& 14.3 (1.75)&  31.5 (97)\\
fe5  & 1.92& 1.67& 5.52& 17.0 (1.02)&  98.6 (42)\\
fe6  & 2.05& 2.00& 2.57& 13.8 (1.00)&  -- (38)\\
fe9  & 2.08& 2.11& 1.22& 15.7 (0.89)& 87.6 (10) \\
fe10 & 2.06& 2.11& 1.51& 15.6 (0.85)& 84.8 (--)\\
fe11 & 2.14& 2.30& 0.00& 20.3& $>99.99$ \\
fe12 & 1.73& 1.39& 16.2& 34.0& $>99.99$ \\
fe13 & 1.76& 1.44& 7.86& 23.8& $>99.99$ \\
fe14 & 1.87& 1.65& 4.46& 17.0 (1.32)& 98.7 (82)\\
\\
fe7  & 1.53& 1.16& 15.1& 28.3& $>99.99$\\
fe8  & 2.40& 2.76& 0.00& 69.8& $>99.99$    \\
\end{tabular}
\label{fits}
\end{table*}

\begin{figure*}
\leavevmode\epsffile{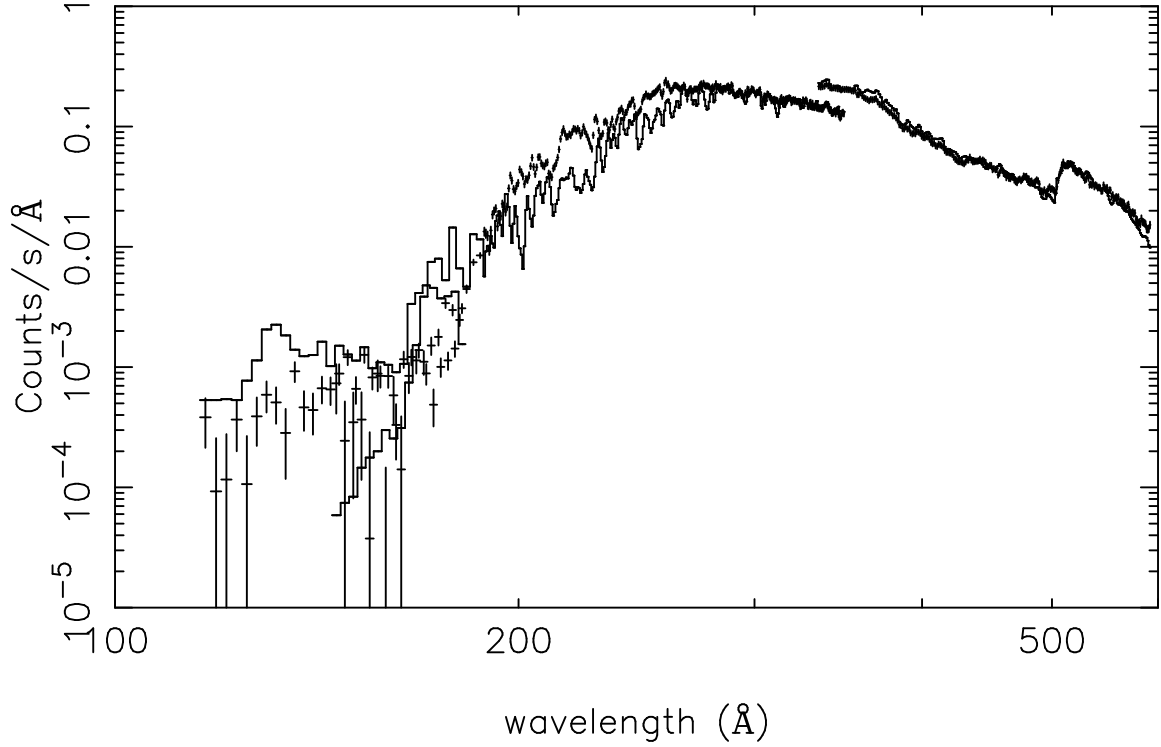}
\caption{ 
Comparison of the complete \eu \ count spectrum of G191$-$B2B with stratified 
model fe1 and best-fit interstellar parameters (\hi $=2.02\times 10^{18}$,
\hei $=1.76\times 10^{17}$, \heii $=7.78\times 10^{17}$,
$M_H=1.3\times 10^{-13}M_\odot$, log g$=7.5$,
\teff $=54000$K). }
\label{f4}
\end{figure*}

\begin{figure*}
\leavevmode\epsffile{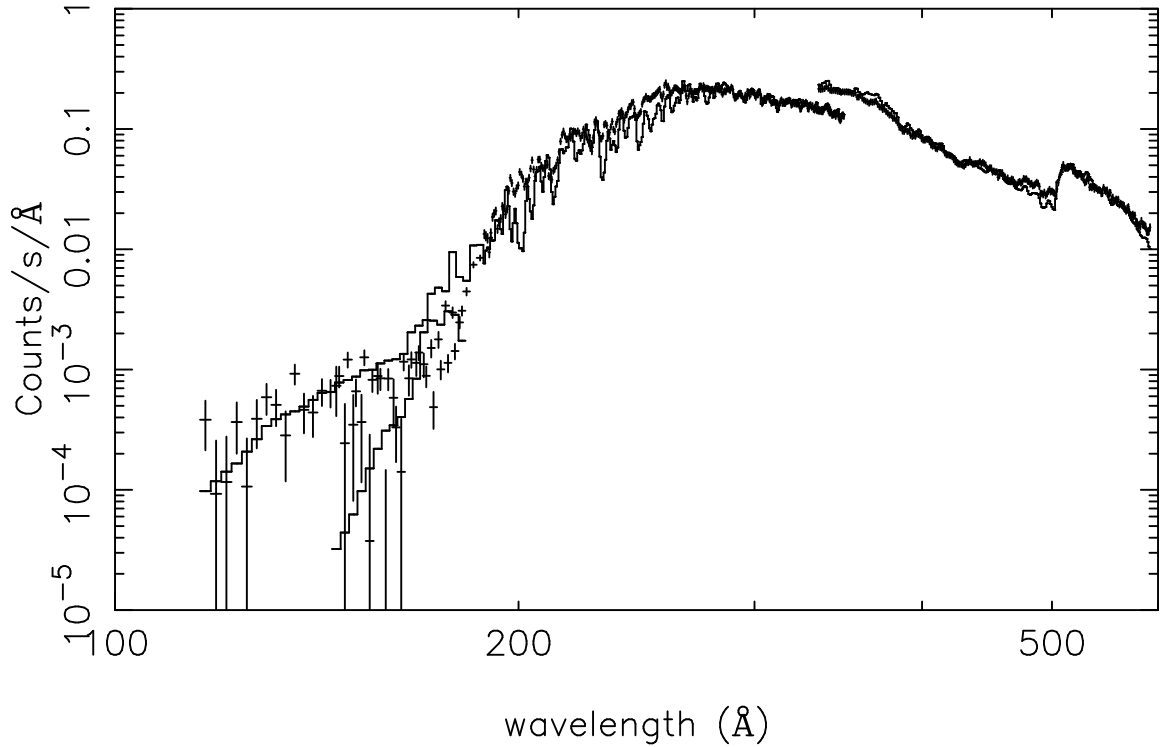}
\caption{
Comparison of the complete \eu \ count spectrum of G191$-$B2B with model fe6 
that gives the optimum match to the entire wavelength range (\hi $=2.05\times 
10^{18}$,
\hei $=2.00\times 10^{17}$, \heii $=2.57\times 10^{17}$,
$M_H=1.3\times 10^{-13}M_\odot$, log g$=7.5$,
\teff $=54000$K). }
\label{f5}
\end{figure*}

\begin{figure*}
\leavevmode\epsffile{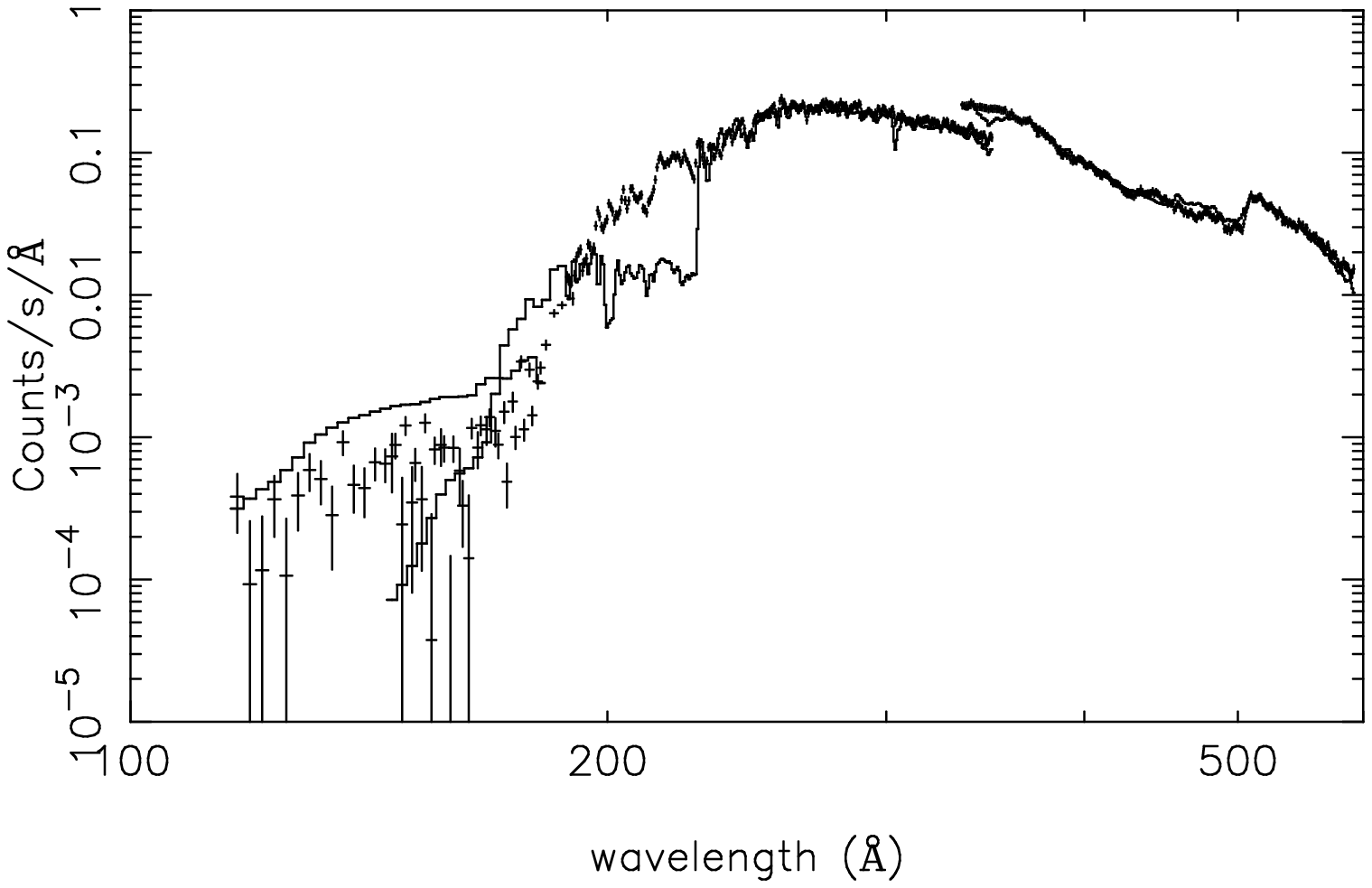}
\caption{
Comparison of the complete \eu \ count spectrum of G191$-$B2B with model fe7 
(\hi $=1.53\times 10^{18}$,
\hei $=1.16\times 10^{17}$, \heii $=15.1\times 10^{17}$,
$M_H=1.3\times 10^{-13}M_\odot$, log g$=7.5$,
\teff $=54000$K). }
\label{f6}
\end{figure*}

Most of the stratified models offer a significant improvement over that 
homogeneous model which gives the best match to the EUV spectrum of G191$-$B2B. 
However, some are rather more successful than others. We examined two different 
types of slab model, first where the upper
layers of the photosphere have the greater Fe abundance
and second with this situation reversed. Model fe1, which falls into the former 
category gives a worse agreement
overall, in comparison with the homogeneous case (figure~\ref{f4}). The predicted 
MW flux clearly falls below the observed level while there is an opposite 
discrepancy in the SW range. Those slab models where the
deeper Fe abundance is greater than that in the outer layers give the best 
agreement in most cases. The very best of these, model fe6, is a good match to 
the observed spectrum throughout the complete EUV range (figure~\ref{f5}). Any 
residual differences are of similar magnitude to those obtained with the 
homogeneous
models when those deal with the more restricted wavelength range above $\approx 
180$\AA . Neither of the two models which have Fe abundance profiles determined 
from the balance of radiative levitation and gravitational settling are in good 
agreement with the observations. For example,
the better of these, fe7, requires a very high \heii \ column density to force 
agreement with the short wavelength flux level. This leaves  an unacceptably 
high flux decrement at and below the 228\AA \ \heii \ Lyman series limit 
(figure~\ref{f6}).

\begin{figure*}
\leavevmode\epsffile{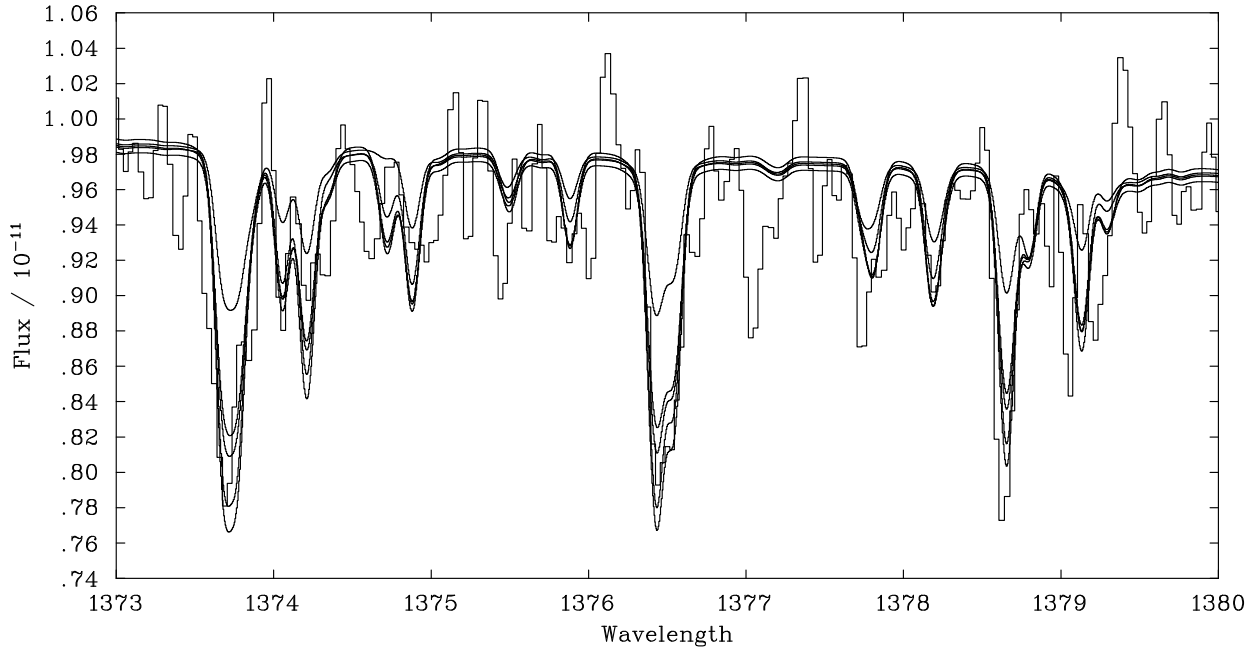}
\caption{Coadded \iu \ echelle spectrum of G191$-$B2B (histogram) spanning the 
range $\approx 1370$ to 1380\AA . Synthetic spectra calculated for models fe4, 
fe3, fe6, fe9 and fe14 (listed in order of decreasing line strength) are shown 
for comparison. 
}
\label{f7}
\end{figure*}

Thus far, the analysis has only addressed the level of agreement between model 
predictions and observations in the EUV spectral range. However, as the Fe 
abundances are
adjusted at different depths to force agreement with the EUV observations, it is 
important to consider the effect this has on the predicted Fe line strengths in 
the far UV.
At this point, we can discard those stratified models
which do not work and limit the analysis to the selected
few that do. Using the F-test, we define this group by assessing which model 
fits have values of $\chi^2_{red}$ which indicate that the probability the fit 
is significantly
worse than the best model (fe6) lies below 99\% \ (see table~\ref{fits}). The 
models included are fe3, fe4, fe5, fe6, fe9, fe10 and fe14. Figure~\ref{f7} 
shows the
a region of the \iu \ NEWSIPS coadded spectrum of G191$-$B2B spanning the range 
$\approx 1370$ to 1380\AA \ and including several of the strongest FeV lines. 
Also shown, in decreasing order of predicted line strength are
synthetic spectra computed for models fe4, fe3, fe6, fe9 and fe14. Models fe4 
and fe3 are a very good match, with
the fe6 and fe9 FeV line strengths being slightly weaker than observed. To assess
quantitatively the level of agreement, we have carried out a further analysis,
fitting the models predictions to the two strongest FeV lines seen in figure~\ref{f7}
(at 1373.8 and 1376.5\AA ). The resulting values of $\chi^2_{red}$ are listed
(in brackets) in table~\ref{fits}, showing that all models
are in good agreement with the data and that none can be excluded on the
basis of the F test. While model fe6 is
the best in the EUV range, fe10, which is not significantly different gives the best 
match to the far-UV FeV lines. 

\section{Discussion}

We have been able to provide the first consistent explanation of the complete 
spectral flux distribution of the hot DA white dwarf G191$-$B2B, including the 
short wavelength EUV region ($<190$\AA ) which has previously been problematic, 
using a grid of atmosphere models with a depth dependent abundance of Fe. 
Potentially, this is an important breakthrough in our understanding of this and 
related DAs with significant photospheric heavy element abundances. However, the 
choice of model structures is somewhat arbitrary, even artificial, in the 
absence of any a priori physical constraints that we might apply. Hence, it is 
necessary to re-examine the justification for this approach, question its 
physical reality and assess the uniqueness of the models before consideration of 
the possible implications of the results.

In comparison with the homogeneous atmosphere models
used in the earlier studies of G191$-$B2B (see figure~\ref{f1}), the stratified 
models described here
seek to suppress the level of the short wavelength ($<190$\AA ) flux compared to 
the longer wavelengths, also steepening the spectral slope towards short 
wavelengths in this region. A number of the models tried were successful,
but there may be other mechanisms that might yield a similar effect. Probably 
the most important question concerns the atomic data included in the models. It 
is well-known that there are large uncertainties in the
continuum and line opacities of the Fe group elements. As the dominant EUV 
opacity source, the Fe data will be the most critical. However, there are two 
arguments againts this possibility. First, it seems unreasonable to suppose that 
any errors should be concentrated in a particular wavelength range. Second, 
where the greatest 
uncertainties lie, below $\approx 160$\AA , there
are very few Fe lines (or of any other species). This might suggest that, in fact, 
there should be less problem in the 
short wavelength region. However, we cannot ignore this potential problem
completely. Bautista (1996) and Bautista and Pradhan (1997) have computed
photoionization cross-section and oscillator strengths for FeIV and FeV, expanding
the Opacity Project database by and appreciable number of new transitions.
Their calculations take account of transitions between states lying below
the first ionization threshold and states lying above. As Bautista and Pradhan
note, these transitions can make an important contribution to the total photoabsorption,
even though they do not appear as resonances in the photoionisation cross-sections.
An alternative explanation for the short wavelength flux
decrement is that extra EUV opacity 
may be present in the form of species, different elements or ionization stages, 
that have not been taken into account in the models. Again, this seems unlikely, 
as
the few detected elements that we have not included (S and P) are only present 
in traces too small to have any noticeable effet in the EUV (Vennes \etal \ 
1996). Any
elements not yet detected will have still lower abundances.

The choice of stratified models is really determined by the level of complexity 
that can be accomodated in the modelling process and by the availability of 
information to determine the abundance depth profile of a given element. The 
two layer slab models calculated, therefore, represent the simplest possible 
development of the idea of a depth dependent abundance. Extending this to deal 
with three layers (models fe1, fe12, fe13, fe14) is a modest increase in 
complexity, adding two further unknowns
(position of the lower layer boundary and the layer abundance) to the three 
required for two layers. In this work, the additional, narrow layer (2-4 depth 
points) 
is used to make a smoother transition between the two main
slabs. Generation of a smooth abundance profile is also a result of the 
diffusion/levitation calculations included
in models fe7 and fe8.

It is rather striking that it is the simplest two region slab models that give 
the best agreement with the data.
Of the three layer models, only fe14, where the intermediate layer is reduced to 
one depth point is in reasonable agreement and the diffusion/levitation models 
are the worst as a group. One conclusion we could draw from this is that the 
transition region for the change in Fe abundance is indeed narrow. In addition, 
the evidence seems to indicate that the Fe abundance is greatest in the deeper 
layers of the atmosphere but with the abundance in
the outer regions being finite and the contrast between
the two values being limited to a factor $\approx 50$. For example, the best fit 
fe6 model (to the EUV data) has a layer abundance ratio of 40, and predicts far 
UV Fe line strengths that are consistent with the observed spectrum.
In comparison, the fe14 model (layer abundance ratio 60), which also gives a 
good match in the EUV, yields far UV
Fe line strengths that are much weaker than observed. This also indicates that 
the abundance of Fe in the outer layer
cannot be less than $\approx 5\times 10^{-7}$. 

It is interesting to note that models fe9 and fe10, which are not significantly
worse than the best model (fe6), have a lower \heii \ column than any of the other
successful models. The resulting He ionization fractions, 37\% \ and 42\% \ respectively,
are much closer to the 27\% \ mean value typical of the local ISM (Barstow
\etal \ 1997), compared with  either the homogeneous 
analysis of Lanz \etal \ (1996; $\approx 80$\% ),
the stratified H/He work of Barstow \& \ Hubeny (1998; $\approx 50$\%) or
the good models considered here (52\% \ for fe6). On the basis of achieving the lowest
He ionization fraction we might favour model fe9 over the others. 

However, we must be cautious in taking any of these results too literally, since there 
are an enormous number of possible, more complex, abundance profiles that we 
have not yet tested and which might give an equally good or better result. 
Furthermore, we have only examined the stratification of Fe. There is no reason 
to assume that
this is the only element that might be stratified. Indeed,
there is good evidence to show that nitrogen is stratified in the slightly 
cooler, less heavy element-rich, DA white
dwarf REJ1032+532 (Holberg \etal \ in preparation). 

Ideally, we should investigate all these possibilities with new calculations but 
there is a major problem in producing the large number of models
needed. We must seek ways of confining the problem.
It may be possible to provide some constraints by measuring abundances using 
lines that are formed at different depths within the envelope. However, this 
approach will be particularly sensitive to any uncertainties in the oscillator 
strengths. Furthermore, as any effects are likely to be quite subtle and the 
range of uncertainty in determining abundances from far UV line strengths
is 
typically a factor 2, it will be necessary to obtain data of considerably higher 
signal-to-noise than that currently available. Even then, as examination of
the EUV and far UV line formation depths shows (figure~\ref{f3}), it will only 
be possible to investigate a narrow region of the photosphere, occupying
1\ dex in 
mass depth. Interestingly, there is much more contrast in line formation depth 
within the EUV and between the EUV and far UV than in the far UV range alone.

If we accept at face value the evidence this work presents, that the Fe in 
G191$-$B2B is stratified in
two main layers with abundances of $\approx 4\times 10^{-5}$ (lower layer) and 
$\approx 1\times 10^{-6}$ (upper layer), it is interesting to explore the 
possible implications. A relative depletion of Fe in the outer layers of the 
envelope may be an indication of ongoing mass-loss in the star. The effects of 
mass-loss in white dwarf atmospheres has hardly been examined, although there is 
evidence for this occuring in the hot DO white dwarf REJ0503$-$289 (Barstow \& \ 
Sion, 1994). A preliminary study of this problem by Chayer \etal 
\ (1993) shows that the outer layers of the envelope will become depleted over 
time.
However, the mass loss rate used in the calculation
($10^{-16}M_\odot$/yr) was sufficient to eliminate
the reservoir of the heavy elements completely within
a few thousand years. On this basis, to see any heavy elements at all, the mass 
loss rate in G191$-$B2B must be considerably lower. There is clearly a need for 
new
radiative levitation calculations coupled with mass-loss
to evaluate this problem properly.

\section{Conclusion}

We have demonstrated that the complete spectrum of G191$-$B2B can be explained 
by a model atmosphere where Fe is stratified, with increasing abundance at 
greater depth. 
The abundance profile appears to be sharply stepped
and may explain the difficulties in matching
observed photospheric abundances, usually obtained by analyses
utilising homogeneous model atmospheres, to the detailed radiative levitation 
predictions. Particularly as the latter are only strictly valid for
regions deeper than where the EUV/far UV lines and continua are formed. Chayer \etal 
\ (1993) show that the outer layers of the envelope will become depleted over 
time if a weak wind is present.
Hence, if found 
to be the only explanation of the observed spectrum, the relative depletion of 
Fe in the outer layers of the atmosphere could be the first evidence for 
radiatively driven mass loss in the star.

In addition, the work presented here may contribute
to the resolution of the
issue of the possible presence of \heii \ along the line of sight
to the star, discussed by
Lanz \etal \ (1996) and Barstow \& \ Hubeny (1998),
and its likely location, in the photosphere or ISM. We find that two of
our best stratified models yield an \heii \ column density and He ionization
fraction closer to the local ISM values than results obtained 
in the earlier studies. However, 
this particular problem will only be completely
solved when a much higher resolution spectrum is obtained, capable
of separating \heii \ lines from those of heavier elements. We
anticipate that the J-PEX spectrometer will provide such data in early 1999
(Bannister \etal \ 1999). Through its ability to study individual lines,
this instrument may also be able to deliver
new information on the stratification of Fe and other elements.

\section*{Acknowledgements}

The work of MAB was supported by PPARC, UK, through an Advanced Fellowship. 
JBH wishes to acknowledge NASA grants NAG 
5-2738 and NAG 5-3472. Data analysis and interpretation were performed using 
NOAO \iraf , NASA HEASARC and Starlink software.

\end{document}